\documentclass[a4paper]{spie}  
\usepackage[english]{babel}
\usepackage[utf8]{inputenc}
\usepackage[]{graphicx}
\usepackage{microtype}
\usepackage{epsfig}
\usepackage{float}
\usepackage{amsmath}
\usepackage{subfig}

\usepackage{natbib}
\bibpunct{(}{)}{;}{a}{}{,}

\graphicspath{{./}}
\usepackage{txfonts}
\usepackage{enumerate}

\providecommand{\supit}[1]{\ensuremath{^{\textit{#1}}}}
\providecommand{\skiplinehalf}{\vspace{1.5ex}\par}

\title{Two Fabry-P\'{e}rots and two calibration units for CARMENES} 

\author{Sebastian Sch\"afer\supit{a}, Eike W. Guenther\supit{b}, Ansgar Reiners\supit{a}, Johannes Winkler\supit{b}, Michael Pluto\supit{b}, J\"org Schiller\supit{b}
\skiplinehalf
\supit{a}Georg August Universit\"at G\"ottingen, Institute for Astrophysics, Friedrich-Hund-Platz 1, 37077 G\"ottingen, Germany \\
\supit{b}Th\"uringer Landessternwarte Tautenburg, Sternwarte 5, 07778 Tautenburg, Germany
}

\begin{document} 
  \maketitle 

\begin{abstract}
The wavelength calibration and nightly drift measurements for  CARMENES (\textbf{C}alar \textbf{A}lto high-\textbf{R}esolution search for \textbf{M} dwarfs with \textbf{E}xoearths with \textbf{N}ear-infrared and optical \textbf{E}chelle \textbf{S}pectrographs) are provided by a combination of hollow cathode lamps and two Fabry-P\'{e}rot units. CARMENES consists of two spectrograph, one for the visible part of the spectrum (520\,-960\,nm) and one for the near infrared (960\,-\,1710\,nm). Each spectrograph has its own calibration unit and its own Fabry-P\'{e}rot. The calibration units are equipped with Th-Ne, U-Ar and U-Ne hollow cathode lamps as well as a flat field lamp. The Fabry-P\'{e}rots are optimized for the wavelength ranges of the spectrographs and use halogen-tungsten lamps as light sources. The Fabry-P\'{e}rots have a free spectral range of 15\,GHz for the visible and 12.2\,GHz for the near infrared which translates to $\sim$17,900 useful emission lines for the visible spectrograph and $\sim$9,700 for the infrared. These lines are used to compute the wavelength solution, and to monitor the instrumental drift during the night. The Fabry-P\'{e}rot units are temperature and pressure stabilized and designed to reach an internal stability of better than 10\,cm/s per night. Here, we present the designs of both Fabry-P\'{e}rot units and the calibration units.
\end{abstract}


\keywords{CARMENES, wavelength calibration, Fabry-P\'{e}rot, Etalon,calibration unit, radial velocity}

\section{INTRODUCTION}
CARMENES is a high resolution instrument installed at the Calar Alto 3.5\,m telescope. Utilizing its two spectrographs it covers the whole wavelength range from 520\,nm to 1710\,nm, with everything below 960\,nm being observed by the VIS arm and everything above being observed by the NIR arm. The main purpose of the instrument is to carry out the CARMENES radial velocity (RV) survey, searching for  Doppler variations in 324 M\,dwarfs caused by their planetary companions (see \cite{2018A&A...612A..49R}). 

In order to find rocky planets around these M-type stars an accurate wavelength calibration and precise measurements of the instrumental drift are needed. CARMENES uses a combination of hollow cathode lamps (HCL) and Fabry-P\'{e}rot interferometer (FP) to create the wavelength solution at the beginning and end of each night. As shown by \cite{Bauer2015}, combining the HCL and FP spectra leads to substantial improvements in the wavelength calibration compared to the traditional approach of using only HCLs. Additionally, the FP provides an excellent nightly drift check when used in simultaneous calibration mode (feeding one fiber with star light and the other with light from the FP). 

In this paper we describe the setup of CARMENES, showing its two calibration units and two FPs (one of each for both spectrographs). In Section~\ref{sec:sim} we present the theoretical calculations for the FP to derive the requirements for pressure and temperature stabilization in order to achieve the goal of an internal FP precission of less than 10\,cm\,s$^{-1}$ over at least one night. Section~\ref{sec:design} showcases the technical design to fulfill these requirements and in Sec.\ref{sec:performance} we demonstrate the performance over the first two years of operation. Finally, in Sec.~\ref{sec:CU} the layout and mechanical design of the two calibration units and the fiber layout are shown.

\section{FABRY-P\'{E}ROT}
\label{sec:FP}


\subsection{SIMULATIONS}
\label{sec:sim}
\begin{figure}
	\centering
		\includegraphics[width=0.5\textwidth]{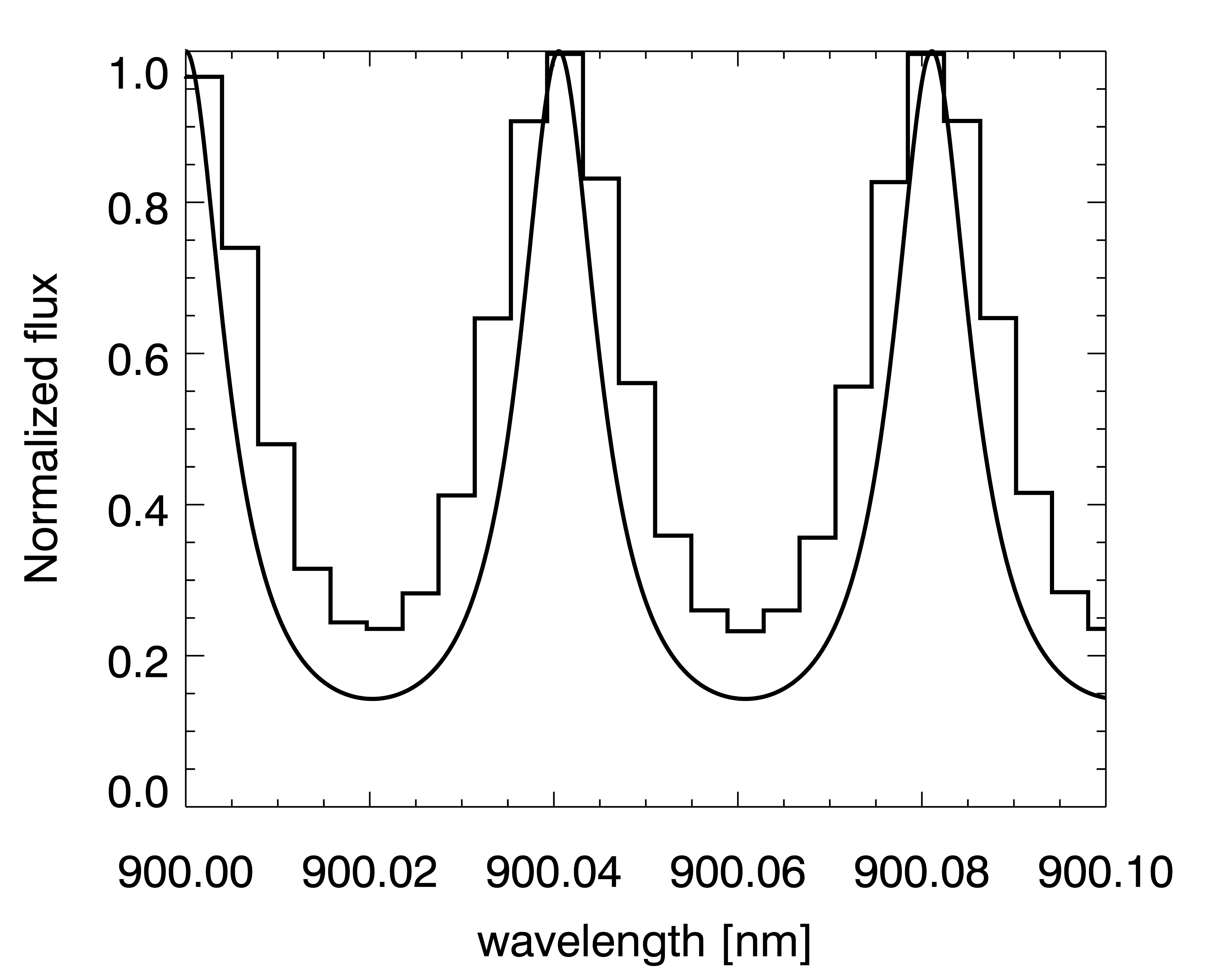}
	\caption{Transmission spectrum of an ideal FP with ${l=1\,\mathrm{cm}}$, ${n=1}$, ${F=14.6}$ and ${\theta=0}$ (smooth line) and the same spectrum sampled with ${R=82,000}$ and 3-pixel sampling.}
	\label{fig:profile_mixed}
\end{figure}
For the ideal case of a plane FP, illuminated by a perfectly collimated beam and assuming the beam to be much smaller than the diameter of the FP, the theoretical transmission function (also called the Airy shape function) can be described as
\begin{equation}
 I = \frac{1}{1 + F \sin^2\left( \frac{\delta}{2} \right)}
 \label{eq:transmisson}
\end{equation}
with the coefficient of finesse $F$ being a function of the reflectivity $R$ of the mirrors used 
\begin{equation}
 F=\frac{4 R}{\left(1-R \right)^2}.
 \label{eq:x1}
\end{equation}
The phase shift $\delta$ is
\begin{equation}
 \delta = \frac{1}{\lambda}4 \pi n l \cos(\theta),
 \label{eq:FP}
\end{equation}
where $\lambda$ corresponds to the wavelength, n is the index of refraction inside the FP, \textit{l} is the distance of the mirrors and $\theta$ the angle between the optical axes of the FP and the light passing through it. The coefficient of finesse $F$ defines the sharpness and the contrast of the transmission peaks whereas $n$, $l$ and $\theta$ are responsible for the position of the peaks and the distance between neighboring peaks, called the free spectral range. For a detailed derivation see for example \cite{Vaughan}. 

Regarding FPs as a tool for wavelength calibration of astronomical spectrographs, these parameters are usually chosen in a way to maximize the spectral information content over the whole wavelength range of the spectrograph used. Butler et al. (1996) showed that the intrinsic Doppler error in any portion of the spectrum due to photon limited errors can be written as 
\begin{equation}
 \sigma_V = \frac{1}{\sqrt{\sum{\left( \frac{dI / dV}{\epsilon}\right)^2}}}
  \label{eq:x2}
\end{equation}
where $dI / dV$ is the local slope of the spectrum and $\epsilon$ is the uncertainty in the residual intensity at that pixel. In order to minimize this error one has to maximize the sum of all $dI / dV$, which, in the case of a FP spectrum, depends on the distance between the FP's mirrors and the resolution of the spectrograph used.

For a high resolution spectrograph like CARMENES the goal is to have the individual peaks unresolved by the instrument (FWHM of the peaks smaller than 3\,x the pixel sampling) but well separated, so the peak-to-peak distance (called free spectral range) is at least equal to 3\,x the pixel sampling. Figure~\ref{fig:profile_mixed} shows the simulated spectrum according to Eq.~\ref{eq:FP} and how a spectrograph like CARMENES would see this spectrum with $R=82,000$ and realistic pixel sampling.

In order to calculate the stability requirements for our external parameters we need to look at their influence on the peak positions. The position of each peak can be derived from the requirement of interference, using Eq.~\ref{eq:transmisson} and Eq.~\ref{eq:FP}:
\begin{equation}
 \frac{2}{\lambda} n l \cos(\theta) = N,
 \label{eq:position}
\end{equation}
where $N$ is an integer numbering each peak. Therefore, changes of the parameters $n$, $l$ and $\theta$ corresponds to a shift of all FP peaks which we can describe as  $\frac{\partial v}{c}$:
\begin{equation}
 \frac{\partial v}{c} \sim \frac{\partial \lambda}{\lambda} \sim \frac{\partial x}{x},
 \label{eq:rv}
\end{equation}
with $x$ being either $n$, $l$ or ${\cos \theta}$. In the case of CARMENES the requirements for the FP ask for a relative RV stability of 10\,cm\,s$^{-1}$ over at least each night. This translates into a required pressure stability of ${\Delta P = 4}$\,x\,${10^{-3}}$\,mbar and a temperature stability of ${\Delta T = 15}$\,mK, see also \cite{schafer2012SPIE}.

The actual spectrum of the FP won't look exactly like shown in Fig.~\ref{fig:profile_mixed}. There are a number of additional factors that can change the shape of the lines. For example one has to take a closer look at the collimated beam. Using fibers to guide the light from a lightsource (e.g. a halogen lamp) to the FP, the light needs to be collimated behind the fiber exit. This can be achieved by using an off-axis parabolic mirror (or a simple lens if one isn't concerned about chromatic aberation). The fiber center is placed into the focal point of the mirror (or lens). However, the light exiting the fiber is not a point source but an extended surface area with a certain near-field and far-field distribution. Assuming a thin lense approximation the deviation from a perfectly collimated beam for each point on the fiberhead can be described by 
\begin{equation}
\theta_{dev} = \arctan \left( \frac{D}{f} \right)
\label{eq:deviation}
\end{equation}
with D being the distance of the ray from the fiber midpoint and f being the effective focal length of the mirror. Depending on the fiber size and light distribution one can now calculate the resulting spectrum for any number of point sources on the fiber head and compute the incoherent sum over all of them to recieve the overall spectrum for a given setup, but that is beyond the scope of this paper.

\subsection{DESIGN AND STABILIZATION}
\label{sec:design}

\begin{figure}[H]
	\centering
		\includegraphics[width=0.7\textwidth]{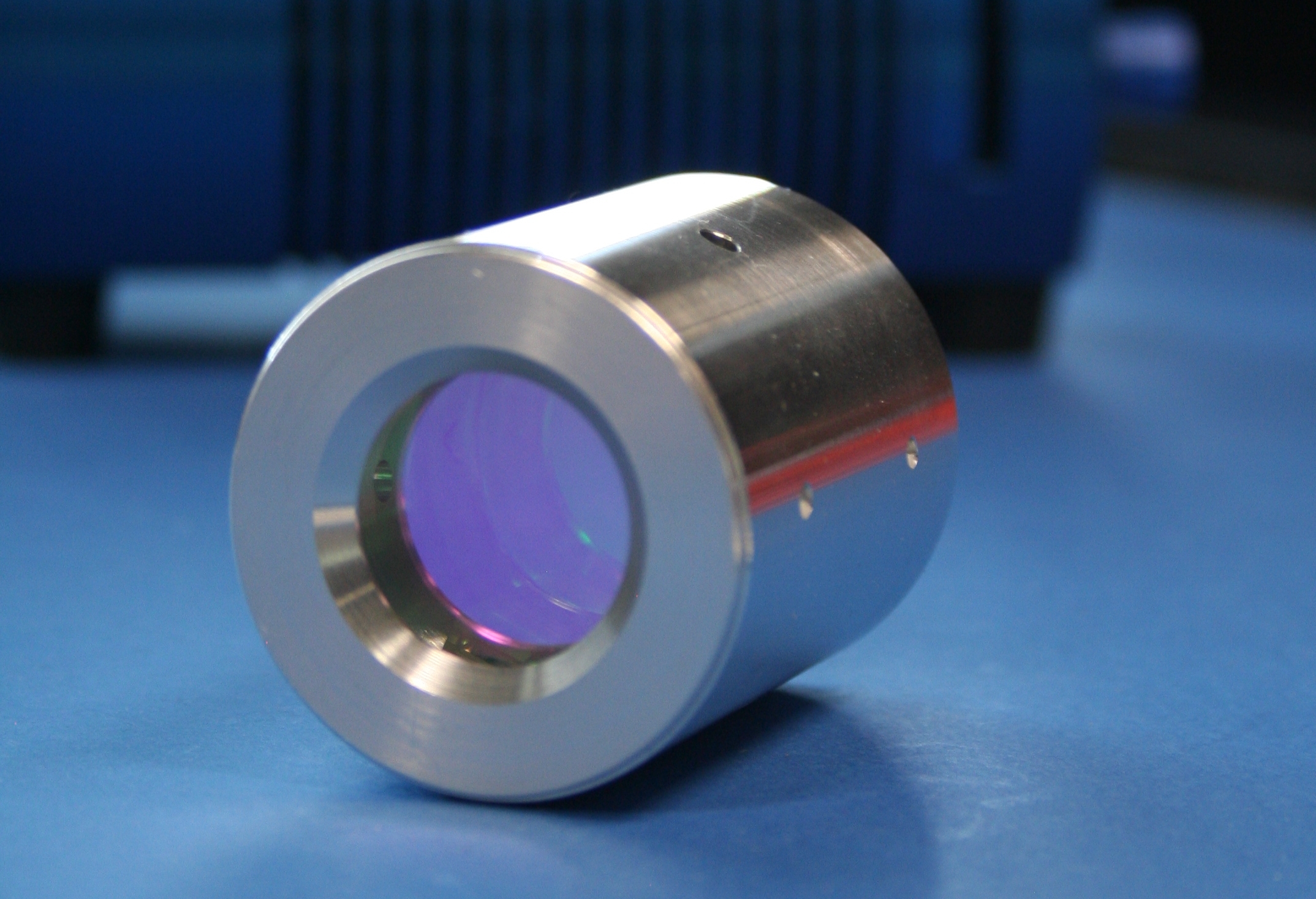}
	\caption{The VIS FP, made by SLS Optics Ltd.}
	\label{fig:fp_foto2}
\end{figure}

CARMENES has two spectrographs, each with its own calibration unit and its own FP unit. Their design is identical with the only difference being the FPs themselves. Both FPs are bought from SLS Optics Ltd. They are air-spaced, soft-coated etalons with wedged mirrors and a coefficient of finesse of $F=26$. The VIS FP is optimized for the wavelength range from 520\,-960\,nm while the near infrared FP is optimized for 960\,-\,1710\,nm using a different coating. They also have a slightly different mirror gap (9.99\,mm for the VIS and 12.3\,mm for the NIR) to achieve a free spectral range of 15\,GHz for the VIS and 12.2\,GHz for the NIR FP respectively, which translates to $\sim$17,900 useful emission lines for the visible spectrograph and $\sim$9,700 for the infrared. Figure\,\ref{fig:fp_foto2} shows the VIS FP before integration into the optical bench.

We use  two LOT halogen lamps with stabilzed  power sources as light sources for the FPs. An additional shutter is placed between the lamp and a Thorlabs RC08SMA-P01 silver coated off axis parabolic mirror that is used to couple the light into the respective input fibers for the two FP units. The shutter is used to maintain a constant count rate on the CCD detector of the spectrograph regardless of exposure time. This is done by frequently opening and closing the shutter such as that the total exposure time of the FPs is always the same for all objects. The input and output fibers are glued into custom made vacuum feedthroughs without any additional connectors. The output fibers are leading to their respective calibration units.

\begin{figure}[H]
	\centering
		\includegraphics[width=0.4\textwidth]{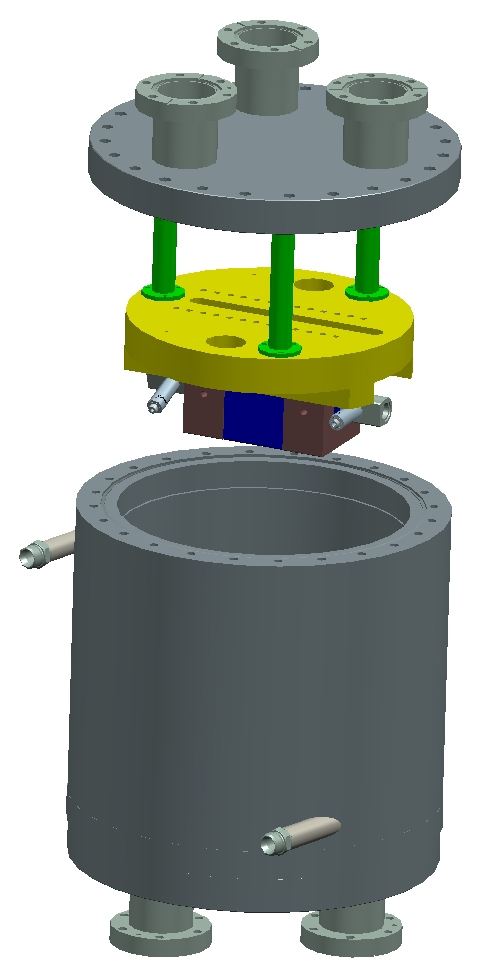}
		\includegraphics[width=0.55\textwidth]{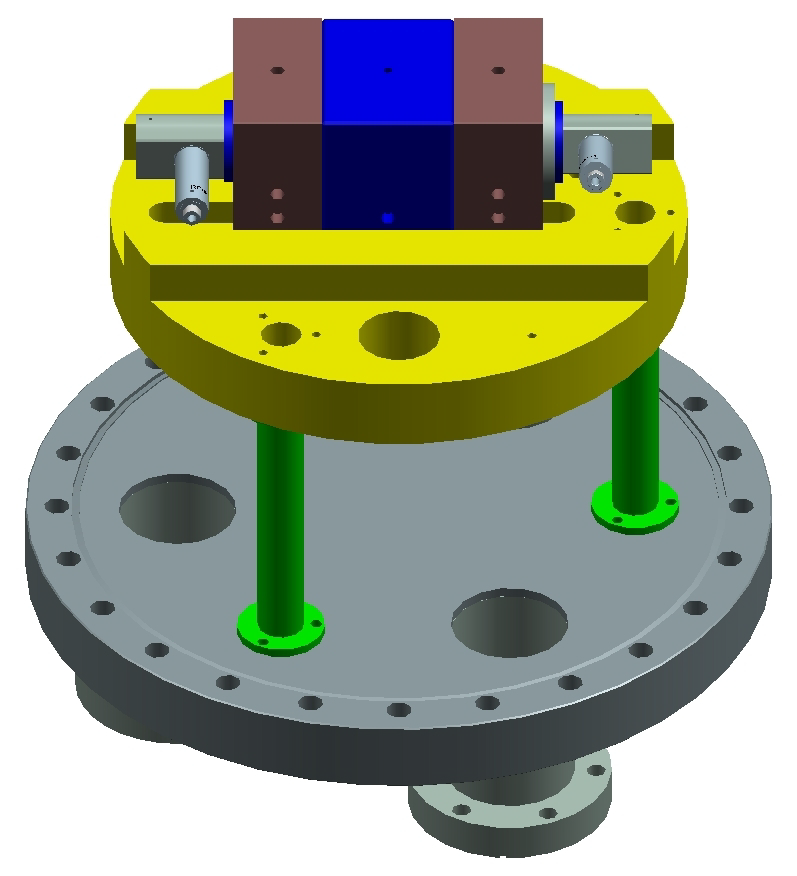}
	\caption{The optical bench (yellow), the Al block holding the FP (blue) and the off axis parabolic mirrors attached (grey). The optical bench is hanging from the top of the vacuum vessel, hold by three rods (green). }
	\label{fig:fp_design}
\end{figure}

Figure\,\ref{fig:fp_design} shows the design of one of the FP units. In order to achieve the required temperature stabilization we designed a double wall vacuum tank. We use a HUBER Ministat 125 compact cooling bath circulation thermostat to keep a silicon oil based thermofluid at a constant ${13^{\circ}}$\,C. Upgraded with the E-Exclusive software, the Ministat 125 provides a temperature resolution of 10\,mK on its in-build PT100 sensor and uses a 5-point calibration to stabilize the temperature. With a heating and cooling-capacity of 1\,kW and 0.3\,kW, respectively, the device is able to  compensate for all external temperature shifts. The pressure pump is pumping the thermofluid into the interlayer at the bottom of the vacuum tank while the suction pump is connected at the upper end of the vessel. The relatively large mass of the vacuum tanks acts as a lowpass filter for any fast temperature changes (timescales less than a minute, e.g. someone entering the room).

The optical bench holding the  FP and the coupling mirrors is mounted at the lid of the vacuum tank, using a tripod design. The FP itself is held inside an aluminum block that is fixed on the optical bench. Thorlabs RC08SMA-P01 silver coated off axis parabolic mirrors are used for both collimating the light from the input fiber and focussing it into the  output fiber after passing through the FP. The mirrors are fixed onto their own aluminum blocks.  

We use a single Pfeiffer HiCube80Eco pumping station for both vacuum tanks. A number of valves allows for different pumping modes or the exchange of one of the three pressure gauges without interrupting operation of the vacuum vessel. The pumping station is configured to minimize the usage of the membrane backing pump to extend its lifetime while keeping the turbopump always active. 

\begin{figure}[H]
	\centering
		\includegraphics[width=0.95\textwidth]{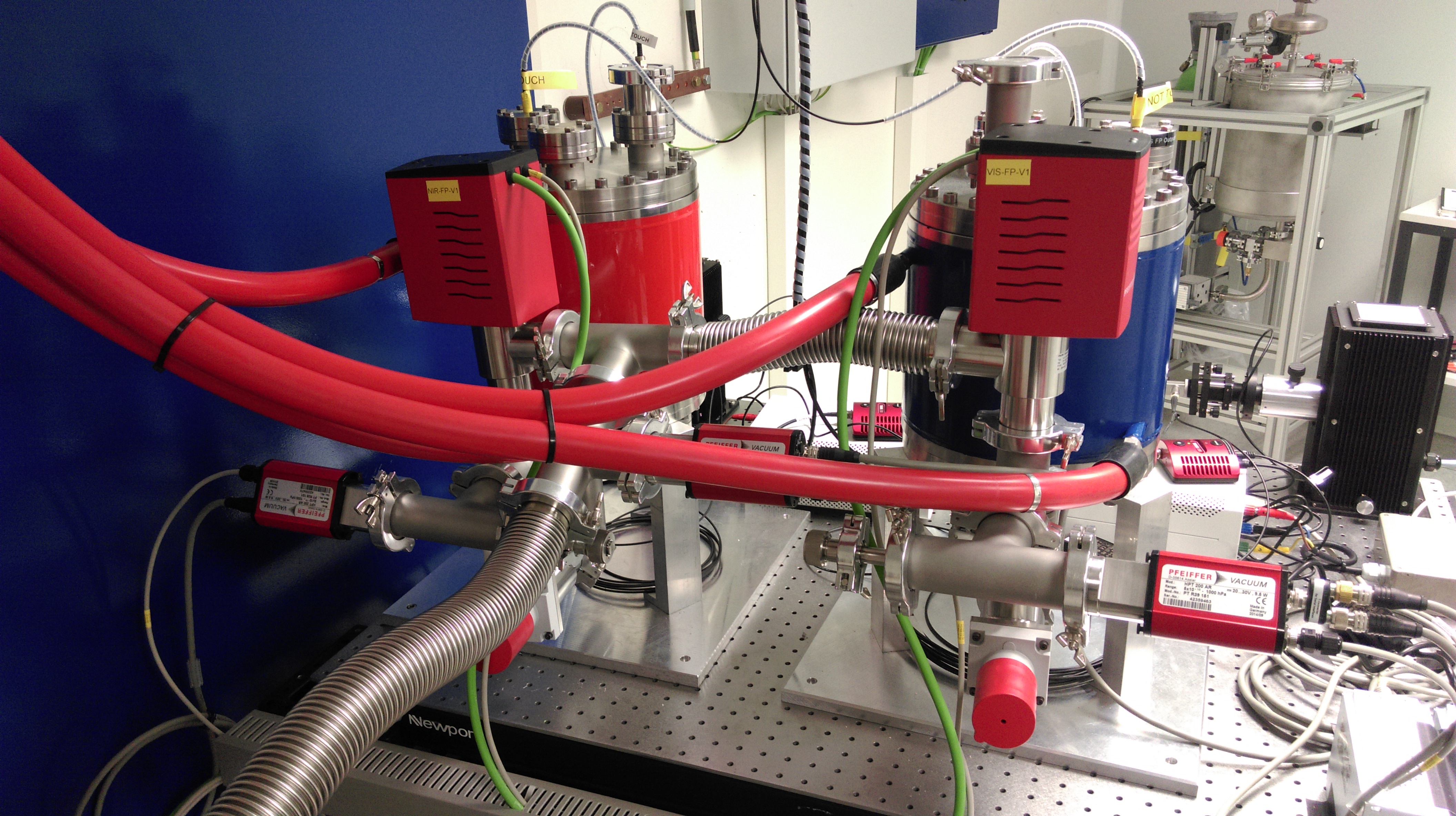}
	\caption{Both FP units (NIR in red and VIS in blue) assembled in the calibration room at Calar Alto with temperature stabilization system (red tubes) and vacuum system attached. The compact cooling bath circulation thermostat and the pumping station are located on a seperate table (not visible). }
	\label{fig:fp_foto1}
\end{figure}

Figure\,\ref{fig:fp_foto1} shows the two FP units (red and blue) assembled in the calibration room of CARMENES. 

\subsection{PERFORMANCE}
\label{sec:performance}
In order to achieve a RV precission of 10\,cm\,s$^{-1}$ the pressure has to stay below 4\,x\,${10^{-3}}$ and the temperature has to be stable within a 15\,mK window around the mean temperature (see Sec.~\ref{sec:sim}). 

\begin{figure}
	\centering
		\includegraphics[width=0.8\textwidth]{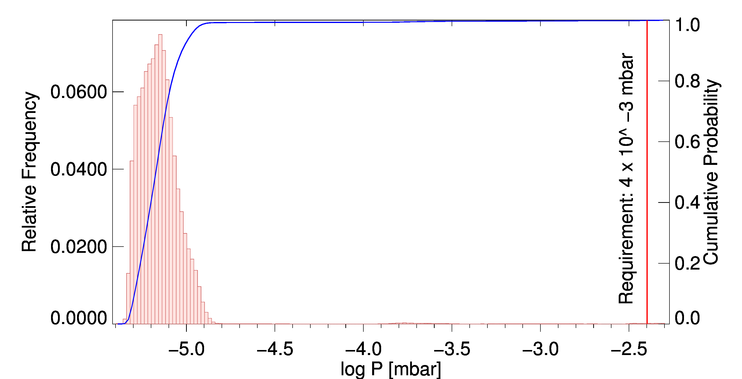}
	\caption{Histogram over all pressure measurements in 2016 and 2017 of the VIS FP in log P [mbar]. The requirement to reach a stability of 10\,cm\,s$^{-1}$ (4\,x\,${10^{-3}}$\,mbar) is indicated by the red vertical line.}
	\label{fig:fp_pressure}
\end{figure}

As a result of the pump station mode the pressure is not stable but fluctuates between a minimum value (reached at the moment the backing pump is turning off) and a maxium value (set by the control loop which is re-activating the backing pump once the pressure reaches the threshold). This cycle is purely caused by the outgassing inside the tanks and its frequency is getting lower with time (about twice per hour at the beginning of 2016 down to once per hour at the end of 2017). Including two years of operation (2016 and 2017) Fig.~\ref{fig:fp_pressure} shows a histogram of the measured pressure inside one of the vacuum tanks. Almost all measurements fall between 5\,x\,${10^{-6}}$ and 1.3\,x\,${10^{-5}}$\,mbar, more than two magnitudes better than the requirement.

The temperature of the thermofluid is measured at the backflow side of the cooling bath circulation thermostat. The thermostat is using its control loop to keep that temperature at the set temperature (13$^\circ$\,C) while the climate control of the room is set to 11$^\circ$\,C. Due to the relatively high mass of the vacuum vessels (over 30\,Kg each) and the additional mass of the optical bench inside we do not expect any high frequency temperature modulations since the mass of the system will effectively damp all external temperature changes. 

\begin{figure}[H]
	\centering
		\includegraphics[width=0.8\textwidth]{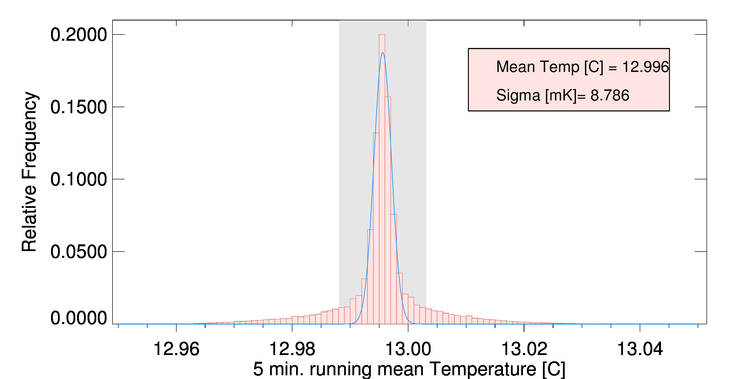}
	\caption{Histogram over all 5-minute running mean temperature measurements in 2016 and 2017 of the VIS FP. The stability requirement is a 15\,mK window around the mean temperature (grey area), 82\% of all measurements are within that region. The blue curve corresponds to a gaussian fit used to find the mean temperature.}
	\label{fig:fp_temp}
\end{figure}
Therefore the requirement for the thermal stability has been set to keep the 5-minute running mean temperature within a 15\,mK window around the mean temperature of the system (which is allowed to differ slightly from the set temperature). Figure~\ref{fig:fp_temp} shows a histogram over two years of said 5-minute running mean temperature measurements (2016 and 2017). We use a gaussian fit to determine the mean temperature of $T=12.996^\circ$C. The standard deviation is 9\,mK, 82\% of all datapoints are within the 15\,mK window. 

Overall we are confident that both temperature and pressure are well within their required ranges in order to achieve a RV precission of 10\,cm\,s$^{-1}$ and we have no indication for any RV-shifts caused by the FP units on any timescale.

\newpage
\section{CALIBRATION UNITS}
\label{sec:CU}
CARMENES has two calibration units, one for the VIS arm and one for the NIR arm. The purpose of the calibration units is to inject light from various source into fibers that transmit the light from these sources to the front end. The light-sources provided are:

\begin{itemize}
 \item Tungsten lamp (for flat-fielding)
 \item Hollow cathode lamps (HCLs)
  \item Fabry-P\'{e}rot
 \end{itemize}

\begin{figure}[H]
	\centering
		\includegraphics[width=0.8\textwidth]{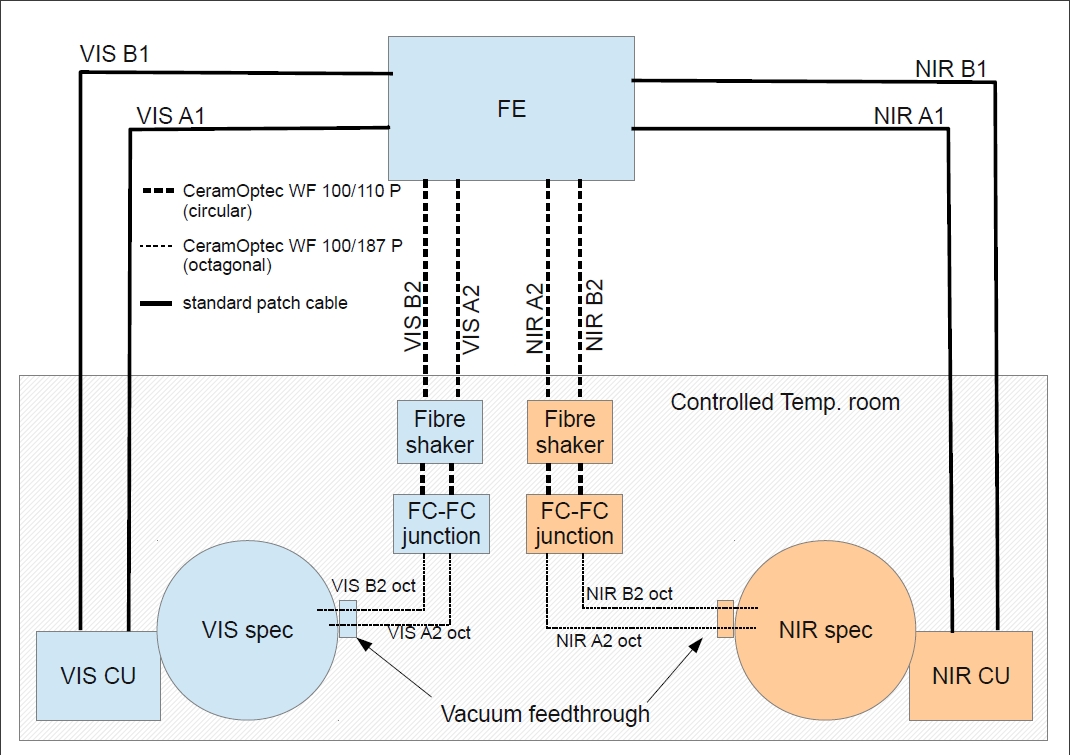}		
	\caption{Schematic layout of fiber-system of CARMENES.}
	\label{fig:calunit_fiberplan}
\end{figure}

Each calibration unit has two fibers leading to the front end. Figure\,\ref{fig:calunit_fiberplan} shows the layout schematically. The fibers connecting the calibration units with the front-end are called A1 and B1 (VIS or NIR). Since the two calibration-units are almost identical, in the following we will simply describe the functionality using 'A1' instead of  'A1 VIS and A1 NIR'. Each unit allows feeding light from a lamp either into none, one, or two optical fibers. The fibers connecting the front-end with the spectrographs are called A2 and B2. 

During observations of stars, starlight is injected into fiber A2. B2 can then either be fed with light from the calibration unit via fiber B1 (simultaneous-calibration mode), or with light from the sky-background (faint-object mode). For daytime calibration (flat-fielding, measuring the wavelength shifts between the fibers), it is possible to feed light either in one of the two fibers (A1 or B1), or in both fibers simultaneously.

Each calibration unit has 7 HCLs for wavelength calibration, and one flat-field lamp. Additionally, there is a port that allows one to feed light from an external, collimated light source which is used as an input for the FPs. CARMENES uses a combination of Uranium-Neon, Thorium-Neon, and Uranium-Argon HCL. 

In order to monitor the health-status of each lamp, we record how many hours, and at which current  each lamp has been used. We also record the voltages of the HCL when they are used. The housing of each HCL lamp is equipped with a Radio-frequency identification chip (RFID) where all this information is stored. After switching on the calibration unit, the RFID units of all lamps are read by the system. Thus, the heath-status of all lamps that are mounted on the Calibration Units is shown. This information is also stored header of the fits-file. 

\begin{figure}
	\centering
		\includegraphics[width=0.7\textwidth]{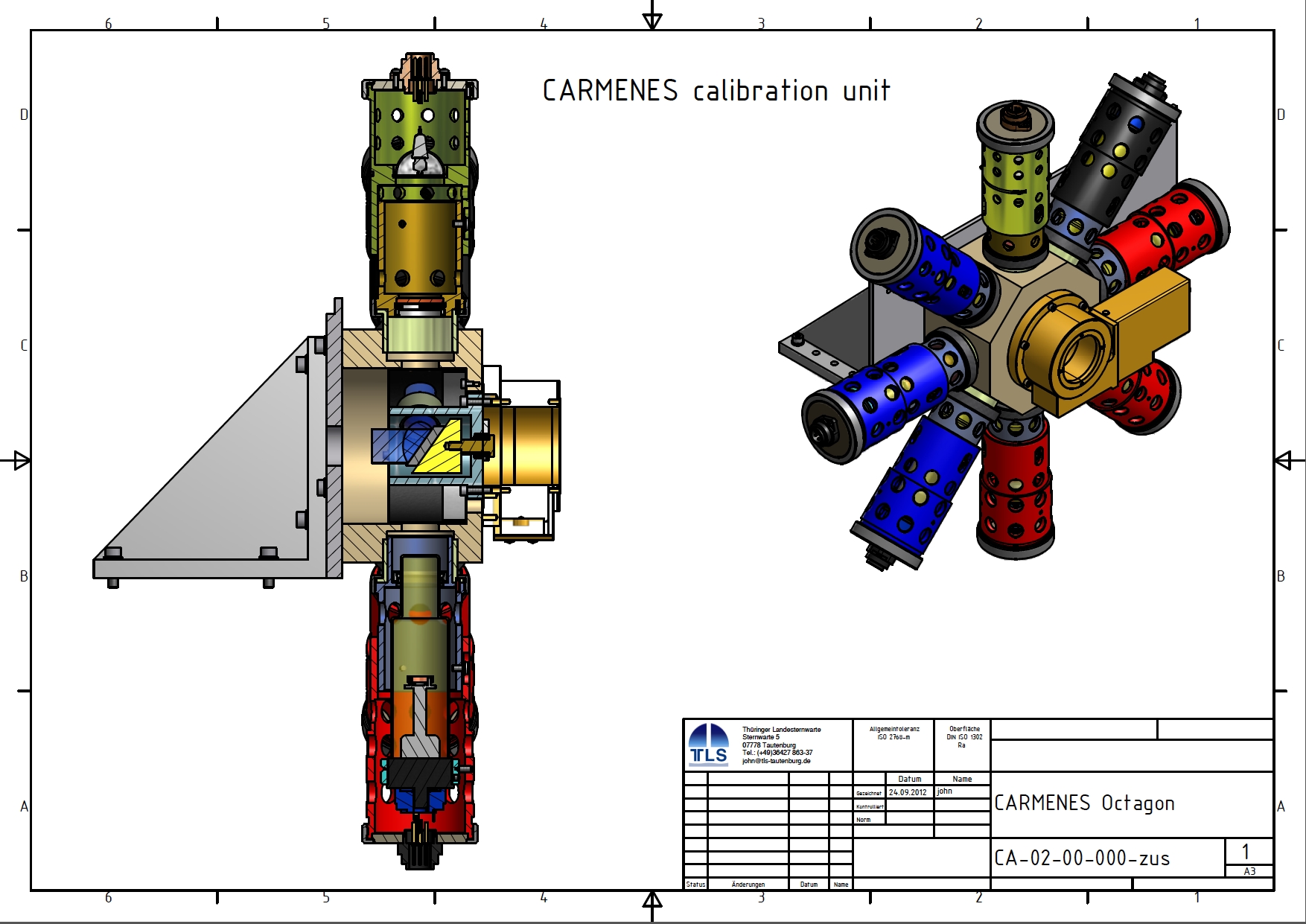}		
	\caption{Drawing of the octagon with the lamps in the rods. At the centre of the octagon is the rotating mirror, which is mounted on the turn-table (yellow).}
	\label{fig:octagon}
\end{figure}

The HCLs and the flat-field lamp are mounted in an octagon. A rotating pick-off mirror in the middle of the octagon is used to pick up the light from the lamp selected (see Fig.~\ref{fig:octagon}). This system is made completely light-tight so that lamps can be switched on for warming-up while a different lamp is being used. Up to 7 HCL can be mounted in each calibration unit. Using the rotating mirror we can switch between lamps within seconds, and we can keep up to four HCL at working temperature. 

\begin{figure}
	\centering
		\includegraphics[width=0.9\textwidth]{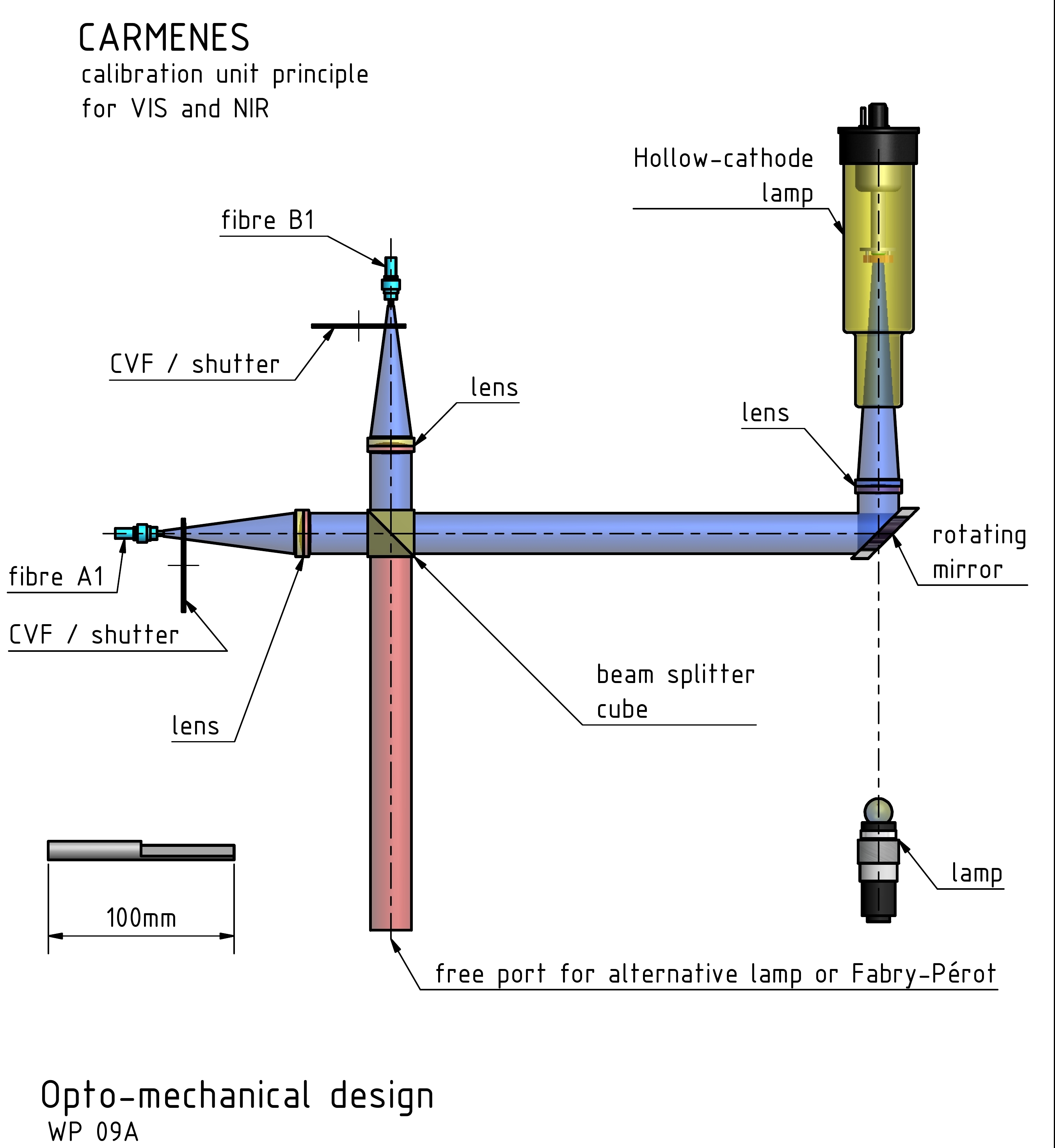}		
	\caption{Schematic layout of the calibration unit.}
	\label{fig:calunit_layout}
\end{figure}

The light from the flat-field and the HCLs is collimated by a lens doublet with a focal length of 125 mm, which sits in the middle of the octagon, next to the mirror that selects the lamps.  A beamsplitter equally splits the light into two  beams A and B. Using a lens of 75 mm focal length the light is injected into the fibers A1 and B1. The only difference between the VIS and NIR units are these lens-systems and the beamsplitter cube which are optimized to their respected wavelength ranges. A schematic layout of the calibration unit is shown in Fig.\,\ref{fig:calunit_layout}. This figure also shows how the light from the FP is injected into the system, and how the rotating shutters allow that light is either injected into one, or both fibers.

\begin{figure}
	\centering
		\includegraphics[width=0.8\textwidth]{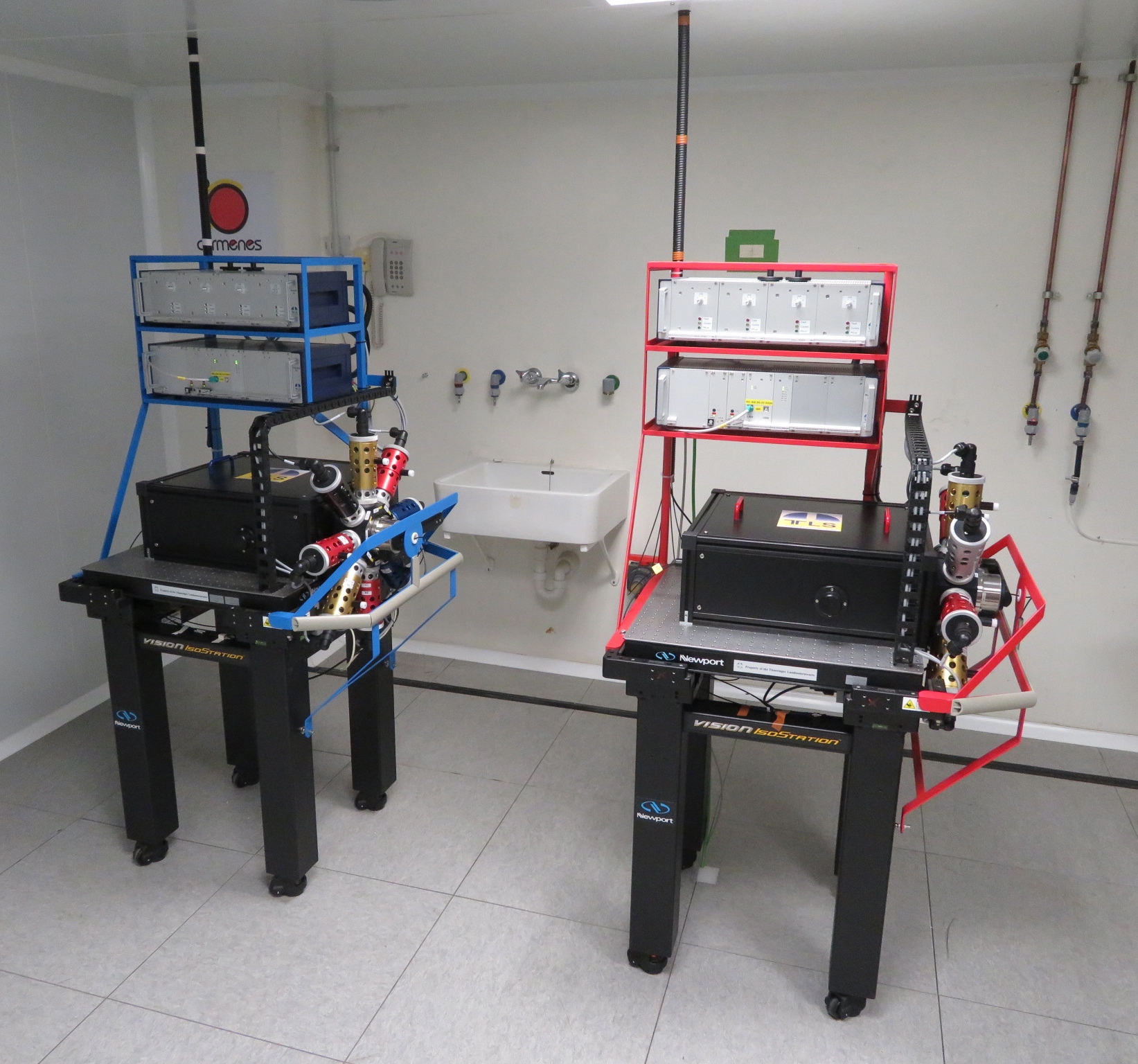}		
	\caption{The two calibrations units on Calar Alto.}
	\label{fig:calunit_foto}
\end{figure}

The current of the HCL can be set to 4 different values: 3, 6, 9, 12 mA. Up to 4 of HCL can be switched on simultaneously but only the light from one lamp is injected into the fiber. Extensive tests have shown that no light from a lamp that is not selected is injected into the fiber, even in an one hour exposure no light from a neighbouring lamp was detected. Each calibration unit is also equipped with two temperature sensors. The temperatures recorded are also stored in the header of the fits-files. Figure~\ref{fig:calunit_foto} shows the two calibration units inside the calibration room on Cala Alto.

\newpage
\section*{ACKNOWLEDGEMENTS}
CARMENES is an instrument for the Centro Astron\'omico Hispano-Alem\'an de  Calar Alto (CAHA, Almer\'{\i}a, Spain). 
  CARMENES is funded by the German Max-Planck-Gesellschaft (MPG), 
  the Spanish Consejo Superior de Investigaciones Cient\'{\i}ficas (CSIC),
  the European Union through FEDER/ERF FICTS-2011-02 funds, 
  and the members of the CARMENES Consortium, 
  with additional contributions by the Spanish Ministry of Economy, 
  the German Science Foundation through the Major Research Instrumentation 
    Programme and DFG Research Unit FOR2544 ``Blue Planets around Red Stars'', 
  the Klaus Tschira Stiftung, 
  the states of Baden-W\"urttemberg, Niedersachsen and Th\"uringen,
  and by the Junta de Andaluc\'{\i}a.
\label{sec:ackn}  


\bibliographystyle{aa}
\bibliography{CARMENES}

\end{document}